\documentclass[twocolumn,showpacs,prl]{revtex4-1}
\usepackage{graphicx}
\usepackage{bm}
\usepackage{amssymb}
\usepackage{amsmath}
\begin{document}

\title{Memory of the Initial Conditions in an Incompletely-Chaotic Quantum System:
Universal Predictions and an Application to Cold Atoms}
\author{V. A. Yurovsky}
\affiliation{School of Chemistry, Tel Aviv University, 69978 Tel Aviv, Israel}
\author{M. Olshanii}
\affiliation{Department of Physics, University of Massachusetts Boston, Boston
MA 02125, USA}
\date{\today}

\begin{abstract}
Two zero-range-interacting atoms in a circular, transversely harmonic waveguide
are used as a test-bench for a quantitative description of the crossover between integrability
and chaos in a quantum system 
with no selection rules.
For such systems we show that the expectation value after relaxation of a
generic observable is given by a linear interpolation between its initial
and thermal expectation values. The variable of this interpolation is universal;
it governs this simple law to cover the whole spectrum of the chaotic
behavior from integrable regime through the well-developed quantum chaos.
The predictions are
confirmed for the waveguide system, where the mode occupations
and the trapping energy were used as the observables of interest; a variety of the initial
states and a full range of the interaction strengths have been tested.
\end{abstract}

\maketitle




Two distinct types of evolution of an isolated dynamical system are 
usually identified: a predictable evolution, strongly correlated with the initial state 
on one hand and a relaxation to the
thermal equilibrium with no memory of the initial state on the other. An ideal gas of
non-interacting atoms is a trivial example of a predictable evolution. In cold
quantum gases, this type of behavior was observed for the single-mode excitations 
in Bose-Einstein condensates
\cite{jin1996,mewes1996}
and a one-dimensional gas of interacting  atoms \cite{kinoshita2006}. Such
systems can be described by integrable models,
i.e.\ models 
that have as many integrals of
motion as there are degrees of freedom. Lifting of integrability leads to a
chaotic motion. 
For ultracold atoms, notable examples of the quantum-chaotic 
motion include both a non-stationary $\delta$-kicked rotor \cite{moore1995,wu2009}
and a stationary billiard \cite{andersen2006}.

Interactions between the 
trapped atoms are the most common cause of integrability lifting
(with the exception of the one-dimensional gas, see
\cite{yurovsky2008b}). 
Even for the case of just two trapped atoms, interactions can already destroy 
integrability \cite{yurovsky2010}. 
For ultracold atoms, the atomic de Brogle wavelength typically substantially exceeds
the interaction range. In this case, the interaction couples every two eigenstates of
non-interacting atoms with approximately the same strength, 
yielding no selection rules interaction can obey. 

In this Letter, we analyze the case of two interacting atoms in a circular,
transversely harmonic waveguide in the multimode regime \cite{yurovsky2010}.
Similar models, the \v{S}eba billiard \cite{seba1990} and a
cylindrically-symmetric harmonic trap with a $\delta$-scatterer 
\cite{idziaszek2005}, were considered 
by other authors as well. These models, as well as the waveguide model, 
are non-integrable and show some 
signatures of a quantum-chaotic behavior as a result. However,
their behavior is only incompletely-chaotic, since the systems demonstrate some
substantial deviations from quantum-chaotic predictions as well (see
\cite{yurovsky2010,stone2010} and the references therein). In these systems, the interaction
strength can be used to tune the ``chaoticity''.

Trajectories of a completely-chaotic classical system fill all the available
phase space in a ``mixing" motion. This leads to a relaxation to the thermal
equilibrium. In quantum systems relaxation to 
the thermal equilibrium---predicted by the Gibbs ensemble---is ensured 
by the
effect of eigenstate thermalization
\cite{deutsch1991,srednicki1994,rigol2008} (see \cite{polkovnikov2010} for a review). 
The relaxation property can
be considered as one of the criteria of chaotic behavior. 
Evolution of an
integrable system preserves the integrals of motion and its 
state after relaxation
is characterized by the generalized Gibbs ensemble \cite{rigol2007}.
It should be noted that completely-chaotic systems are as rare as integrable
systems.

In classical systems, the KAM theory predicts a continuous crossover
between the regular and chaotic regimes, controlled by the 
system parameters. 
For quantum systems, a smooth 
breakdown of thermalization
was demonstrated 
for hard-core bosons on a finite lattice 
\cite{rigol2009}. In this Letter, we offer an approximate
analytic prediction for the
state after relaxation 
of general quantum systems 
with no selection rules (e.g., models \cite{yurovsky2010,seba1990,idziaszek2005} 
and an embedded random matrix model \cite{Olshanii2009}).
The resulting expression smoothly covers the space between the integrable and
quantum-chaotic regimes as a function of an universal parameter which is the same
for all observables and insensitive to the initial state.  
Unlike the predictions for completely-chaotic \cite{rigol2008} and integrable
\cite{rigol2007} systems, the equilibrium-state universality  is not governed 
by the conserved quantities 
alone
but rather by some universal traits in how the 
integrals of motion can be broken. 
The prediction is tested against the exact analytic results for the waveguide system.
This system seems to be experimentally realizable, as
individually-trapped pairs of ultracold atoms are already used in recent experiments
(see \cite{danzl2010} and references therein).




Consider an integrable (IS) and a non-integrable (NS) systems with Hamiltonians
$\hat{H_0}$  and 
$\hat{H}=\hat{H_0}+\hat{V}$, eigenstates $| \vec{n} \rangle$ 
($\vec{n}$ is an appropriate set of quantum numbers)
and $| \alpha\rangle$ (energy ordered), and
eigenenergies $E_{\vec{n}}$ and $E_{\alpha}$, respectively.
The Schr\"odinger equation for NS readily gives 
\begin{equation}
| \alpha \rangle=\sum_{\vec{n}}\frac
{| \vec{n} \rangle \langle\vec{n} |\hat{V}| \alpha \rangle}
{E_{\alpha}-E_{\vec{n}}} .
\label{alpha_gen}
\end{equation}  Given a
non-equilibrium initial state $\hat{\rho}_{\text{in}}\equiv\hat{\rho}(t=0)$,
where  $\hat{\rho}(t)$ is the NS density matrix, the
expectation value of a generic observable $\hat{A}$ of NS relaxes to the infinite time
average (see \cite{rigol2008})
\begin{equation}
A_{\text{rel}} \equiv \lim_{T\to\infty} \frac{1}{T}
\int\limits_{0}^{T}d t \text{Tr}\left(\hat{A}\hat{\rho}(t)\right) 
=\sum_{\alpha} \langle \alpha | \hat{\rho}_{\text{in}}|\alpha \rangle 
\langle \alpha | \hat{A} | \alpha \rangle.
\label{InfTimeAv}
\end{equation}

First, consider the case when (a) initial density matrix is diagonal in the IS
basis, 
$\langle\vec{n} |\hat{\rho}_{\text{in}}| \vec{n}' \rangle=
\rho^{\text{in}}_{\vec{n}} \delta_{\vec{n}\vec{n}'}$,
and (b) the observable is an integral of motion of the IS, 
$\langle\vec{n} |\hat{A}| \vec{n}' \rangle=
A_{\vec{n}} \delta_{\vec{n}\vec{n}'}$.
Using the  partial fraction decomposition, Eqs.\ (\ref{alpha_gen}) and (\ref{InfTimeAv})
can be transformed, with no approximations, to
\begin{subequations}
\begin{eqnarray}
 A_{\text{rel}}=\sum_{\vec{n}}
A_{\vec{n}}\eta_4(E_{\vec{n}})\rho^{\text{in}}_{\vec{n}}+
\sum_{\vec{n}\ne \vec{n}'}
A_{\vec{n}}F_{\vec{n} \vec{n}'}\rho^{\text{in}}_{\vec{n}'} ,
\label{ArelaxF}
\\
F_{\vec{n} \vec{n}'}=
\frac
{\eta_{\vec{n}\vec{n}'\vec{n}\vec{n}'}^{(2)}(E_{\vec{n}})
+\eta_{\vec{n}\vec{n}'\vec{n}\vec{n}'}^{(2)}(E_{\vec{n}'})}
{(E_{\vec{n}}-E_{\vec{n}'})^{2}}
\nonumber
\\
+2\frac
{\eta_{\vec{n}\vec{n}'\vec{n}\vec{n}'}^{(1)}(E_{\vec{n}})
-\eta_{\vec{n}\vec{n}'\vec{n}\vec{n}'}^{(1)}(E_{\vec{n}'})}
{(E_{\vec{n}'}-E_{\vec{n}})^{3}}
\label{FEnEnp}
\\
\eta_{\vec{n}\vec{n}'\vec{n}''\vec{n}'''}^{(j)}(E)=\sum_{\alpha} \frac{ 
V_{\vec{n} \alpha}V_{\vec{n}' \alpha}V_{\alpha \vec{n}''}V_{\alpha \vec{n}'''}}
{(E_{\alpha}-E)^j}.
\label{hateta}
\end{eqnarray}
\end{subequations}
Here
\begin{equation}
\eta_4(E_{\vec{n}})=\eta_{\vec{n}\vec{n}\vec{n}\vec{n}}^{(4)}(E_{\vec{n}})
\equiv \sum_{\alpha}|\langle\vec{n}|\alpha\rangle|^4  
\label{IPR}
\end{equation}
is the inverse participation ratio (IPR) \cite{georgeot1997}.
Its inverse estimates the number of the NS eigenstates 
the IS one consists of. 
IPR can be considered as a measure of
``chaoticity'' of the system, since it varies from 
$1$ for IS to $\eta_4(E_{\vec{n}})\ll 1$ for a completely-chaotic
system.

Diagonal matrix elements $A_{\vec{n}}$ of a typical observable $\hat{A}$ 
can be decomposed into a sum of a smooth function of the state energy 
$E_{\vec{n}}$ and fluctuations around it. 
One can introduce  a macroscopic energy scale 
$\Delta_{\text{MS}}$---an energy over which the smooth part does not
change substantially. Now, assume that the function $F_{\vec{n} \vec{n}'}$ decays at 
the energy distances
$|E_{\vec{n}}-E_{\vec{n}'}|$
less than $\Delta_{\text{MS}}$.
Let us also introduce a set of intervals $[{\cal E}_i,{\cal E}_{i+1}]$ and
the function
 $B_i(\vec{n})=\sum_{\vec{n}',\vec{n}'\ne
\vec{n},{\cal E}_i<E_{\vec{n}'}<{\cal E}_{i+1}}
F_{\vec{n} \vec{n}'}\rho^{\text{in}}_{\vec{n}'}$, 
which is localized in
some energy window $E_{\vec{n}}\in{\cal W}_{i}$.
For a sufficiently small interval $[{\cal E}_i,{\cal E}_{i+1}]$ the 
energy window ${\cal W}_i$ will become smaller than or comparable to $\Delta_{\text{MS}}$.

We consider here a system with no selection rules, i.e.\ we assume that $V_{\vec{n} \vec{n}'}$
do not show any systematic dependence on the differences between the quantum
numbers $\vec{n}$ and $\vec{n}'$. Two distinct classes of such interactions are the
zero-range interactions considered below and the random-matrix
interactions (see \cite{Olshanii2009}).
In the absence of the selection rules, the only systematic dependence between eigenstates 
$\vec{n}$ and $\vec{n}'$, coupled by the function $F_{\vec{n} \vec{n}'}$ can be 
due to the energy denominators in Eq.\ (\ref{FEnEnp}) selecting energy-neighboring states.
However, in the generic case, different degrees of freedom of the IS have 
incommensurate frequencies. 
As a result,
the quantum numbers of highly-excited energy-neighboring states will be 
mutually uncorrelated and $B_i(\vec{n})$ 
is indiscriminate with respect to the
quantum number values available in the window ${\cal W}_{i}$, even if 
IS eigenstates $\vec{n}'$ in the initial state
are specially selected according to their quantum numbers.
Consequently, the sequences 
$A_{\vec{n}}$ and $B_i(\vec{n})$ become uncorrelated.  
Mathematically this
means that their correlation function is equal to the product of their mean values, 
$\sum_{\vec{n}} A_{\vec{n}}B_i(\vec{n})/{\cal N}_{i}
\approx A_{\text{MC}}(({\cal E}_i+{\cal E}_{i+1})/2) \sum_{\vec{n}} B_i(\vec{n})/{\cal N}_{i}$.
Provided a statistically sufficient number of IS 
eigenstates ${\cal N}_{i}$ in the window ${\cal W}_{i}$,
\begin{equation}
A_{\text{MC}}(({\cal E}_i+{\cal E}_{i+1})/2)=\frac{1}{{\cal N}_{i}}
\sum_{\vec{n},E_{\vec{n}}\in{\cal W}_{i}}A_{\vec{n}}
\end{equation}
expresses the definition of the microcanonical expectation value.
Equation (\ref{ArelaxF}) is reduced then to
\begin{equation}
 A_{\text{rel}}\approx\sum_{\vec{n}}\left[A_{\vec{n}}\eta_4(E_{\vec{n}})+
A_{\text{MC}}(E_{\vec{n}})(1-\eta_4(E_{\vec{n}}))\right]\rho^{\text{in}}_{\vec{n}} .
\label{ArelaxEn}
\end{equation}
[Here orthogonality and completeness of the basis set $| \alpha \rangle$ 
are used and $A_{\text{MC}}(({\cal E}_i+{\cal E}_{i+1})/2)$ is approximated by
$A_{\text{MC}}(E_{\vec{n}})$ .]
Further, we can approximate IPR by its average value over the initial state
$\bar{\eta}=\sum_{\vec{n}}\eta_4(E_{\vec{n}})\rho^{\text{in}}_{\vec{n}}$, getting
\begin{eqnarray}
 A_{\text{rel}}\approx\bar{\eta} A_{\text{in}}+(1-\bar{\eta}) A_{\text{therm}}
\label{ArelaxAv}
\\
A_{\text{in}}=\sum_{\vec{n}}A_{\vec{n}}\rho^{\text{in}}_{\vec{n}},
\quad
A_{\text{therm}}=\sum_{\vec{n}}A_{\text{MC}}(E_{\vec{n}})\rho^{\text{in}}_{\vec{
n}} .
\label{AInTherm}
\end{eqnarray}
This expression has a clear physical meaning. An initial state 
populates $\sim\bar{\eta}^{-1}$ states of NS per each IS state $| \vec{n} \rangle$ 
contained in the initial state. Since weight of each of the NS state in 
$| \vec{n} \rangle$ is $\sim\bar{\eta}$ and vice versa, the contribution of 
$| \vec{n} \rangle$ into the equilibrium state will be proportional to $\bar{\eta}$. 
Other IS states, contained in the populated NS states, give a thermal contribution 
of a  weight $1-\bar{\eta}$.
In the case of IS, $\bar{\eta}=1$ and the system keeps initial
expectation value of the observable $A_{\text{in}}$; on the other hand, for
completely-chaotic systems, where $\bar{\eta}\ll 1$, it relaxes to the microcanonical
expectation value, averaged over the initial state, i.e.\ to $A_{\text{therm}}$. 
If $\Delta_{\text{MS}}$ is greater then the energy width of the initial state with the energy $E$,
$A_{\text{therm}}\approx A_{\text{MC}}(E)$ and the memory of initial conditions
is given by the first term in Eq.\ (\ref{ArelaxAv}) only. However, for broad
initial states, $A_{\text{therm}}$ depends on $\hat{\rho}_{\text{in}}$ as well;
in this case, some memory of initial conditions will be retained, even in the 
completely-chaotic regime. 

Since averages over intervals $[{\cal E}_i,{\cal E}_{i+1}]$ fluctuate independently, 
the accuracy of Eq.\ (\ref{ArelaxEn}) is determined by the total number of IS eigenstates
${\cal N}_{\text{A}}$ which are effectively involved in the  summation over $\vec{n}$
in the second term of Eq.\ (\ref{ArelaxF}). Since each eigenstate  $\vec{n}'$ in the 
initial state provides $\sim\bar{\eta}^{-1}(1-\bar{\eta}^{-1})$ eigenstates  $\vec{n}$
to the 
state after relaxation,
\begin{equation}
{\cal N}_{\text{A}}\sim\frac{{\cal N}_{i}A_{\text{MC}}(E)^2}
{\sum_{\vec{n},E_{\vec{n}}\in{\cal W}_{i}}A_{\vec{n}}^2}
\bar{\eta}^{-1}(1-\bar{\eta}^{-1})
\left[\sum_{\vec{n}'}\left(\rho^{\text{in}}_{\vec{n}'}\right)^2\right]^{-1} .
\end{equation}  
Here the last factor  gives number of eigenstates in the initial state 
and the first one estimates part of eigenstates selected by the observable 
$\hat{A}$ (it is approximately independent of the interval ${\cal W}_{i}$
containing $E$).
The applicability criterion would be ${\cal N}_{\text{A}}\gg 1$.




Below, we verify the prediction (\ref{ArelaxAv}) and (\ref{AInTherm}) 
using an example of two ultracold trapped atoms \cite{yurovsky2010}. 
The interaction is approximated by the zero-range Fermi-Huang pseudopotential 
$\hat{V} = (2\pi\hbar^2 a_{s}/ \mu)\delta_{3}({\bm r}) (\partial/\partial r)(r\,\cdot )$ 
(its applicability to ultracold
collisions was widely confirmed, see e.g. \cite{chin2010}).  Here $a_{s}$ is the
three-dimensional $s$-wave scattering length, $\mu$ is the reduced mass of the
colliding atoms, and ${\bm r}$ is the relative coordinate.
The atoms are trapped in a cylindrical harmonic potential with the frequency
$\omega_{\perp}$. The ring geometry of the waveguide imposes period-$L$ boundary
conditions along the potential axis. This model allows separation of the
center-of-mass motion, leaving a system with two degrees of freedom, the axial
$z$ and radial $\rho$ relative coordinates. The eigenstates of non-interacting
IS $| n l \rangle$ are products of the axially-symmetric wavefunction $| n
\rangle$ of two-dimensional harmonic oscillator and a symmetric plane wave with
the momentum $2\pi l\hbar/L$ (states of other symmetry are not coupled by the
zero-range interaction). 
For the zero-range potential, Eq.\ (\ref{alpha_gen}) expresses 
the eigenstate $| \alpha \rangle$ in an explicit form  
\cite{yurovsky2010,seba1990,idziaszek2005}, which involves only a few states 
$| \vec{n} \rangle$ with closest energy. Such systems 
do not approach the regime of a complete chaos, even
for strong interaction; accordingly, their IPR remains relatively large in this regime. 
The present waveguide system retains $\bar{\eta}\gtrsim 0.39$ even when it approaches 
the maximally-chaotic regime (see \cite{yurovsky2010}) at $a_s>0.1a_{\perp}$,
where $a_{\perp }=( \hbar/\mu \omega_{\perp }) ^{1/2}$ is the transverse
oscillator range.
The aspect ratio is chosen as a large transcendental number
$L/a_{\perp}=\pi^{7/2}(1+\sqrt{5})^{1/2}\approx 99$, where the system behavior
appears to be  more chaotic.

Having in mind 
the further comparison with the case of non-diagonal initial state, we
consider the diagonal matrix
$\rho^{\text{in}}_{nl}=|\langle\psi_{\text{in}}| n l \rangle|^2$ with
\begin{equation}
 \langle z,\rho |\psi_{\text{in}}\rangle 
\propto
\cos{\pi \zeta \over \delta}\theta \left(
 {\delta\over 2}-|\zeta |\right) 
\exp\left(-\frac{\kappa\rho^{2}}{a_{\perp}^{2}}\right)
\label{psi_in_dk}
\end{equation} 
and $\zeta =  z/L-1/2$ (see \cite{yurovsky2010}). 
\begin{figure}
\includegraphics*[width=3.4in]{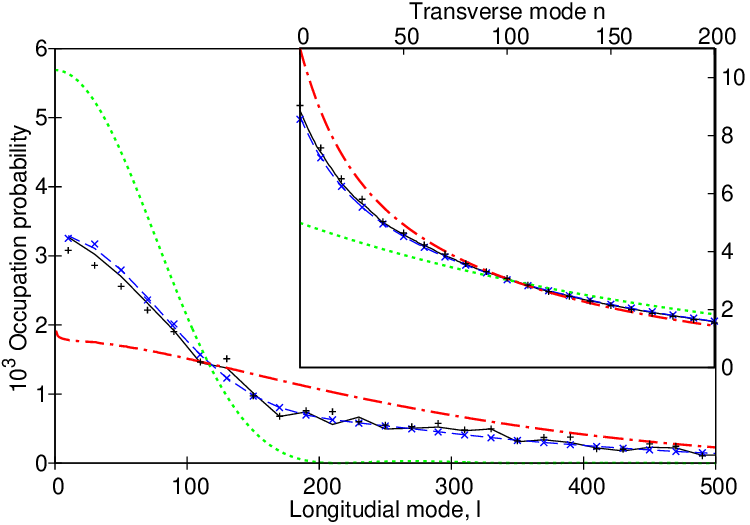}
\caption
{(Color online) Distributions over the transverse
(insert) and longitudinal modes of the integrable system. The infinite-time
average (\protect\ref{InfTimeAv}) [black solid line] is compared to predictions
(\protect\ref{ArelaxEn}) with state-dependent IPR [crosses] and
(\protect\ref{ArelaxAv}) with averaged IPR $\bar{\eta}=0.39$ (blue dashed line)
in the strong-interaction regime $a_{s} = 10^{6} a_{\perp}$ for the initial 
state (\protect\ref{psi_in_dk}) with $\kappa=400$ and
$\delta\approx7\times10^{-3}$.  
The dotted (green) and dot-dashed (red) lines
show the initial distributions and the averaged microcanonical predictions
(\protect\ref{AInTherm}). 
The infinite-time average for the non-diagonal initial density matrix is shown
with pluses.
}
\label{Fig_dist}
\end{figure}

Figure \ref{Fig_dist} demonstrates a good agreement between the predictions
(\ref{ArelaxEn}) and the infinite time average (\ref{InfTimeAv}) for
distributions over transverse and longitudinal modes described by operators 
$|n \rangle \langle n|$ and 
$\frac{1}{20}\sum_{l'=l-9}^{l+10} | l' \rangle \langle l'|$, respectively. 
It also shows that predictions for constant IPR
(\ref{ArelaxAv}) and state-dependent IPR (\ref{ArelaxEn}) almost coincide.




Consider now the more general case when the initial density matrix has
non-diagonal elements in the IS basis. Infinite-time average can be expressed
 like Eq.\ (\ref{ArelaxF}). The contributions of alternating-sign sums 
$\eta_{\vec{n}\vec{n}'\vec{n}''\vec{n}'''}^{(j)}(E)$ 
with odd $j$ can be neglected.
Neglecting also alternating-sign terms with the odd powers of energy differences 
of the involved IS states, we obtain the correction to Eq.\ (\ref{ArelaxF})
\begin{equation}
 A_{\text{nd}\rho}=-2\mathrm{Re} \sum_{\vec{n}\ne \vec{n}'} A_{\vec{n}} 
\eta_{\vec{n}'\vec{n}\vec{n}\vec{n}}^{(2)}(E_{\vec{n}})
\frac{\langle\vec{n} |\hat{\rho}_{\text{in}}| \vec{n}' \rangle}
{(E_{\vec{n}}-E_{\vec{n}'})^{2}}
\label{Andrho}
\end{equation}
This correction is small if the non-diagonal matrix elements of
$\hat{\rho}_{\text{in}}$ have arbitrary phase or if the non-equilibrium initial
state does not contain energy-neighboring modes $\vec{n}$ and $\vec{n}'$ (note,
that uniform occupation of all modes corresponds to the thermal equilibrium).
The initial state of the form 
$\langle n l |\hat{\rho}_{\text{in}}| n' l' \rangle=
\langle n l|\psi_{\text{in}}\rangle \langle \psi_{\text{in}}| n' l'
\rangle$,
used in Fig.\ \ref{Fig_dist}, could lead to large corrections since the
non-diagonal terms are comparable to the diagonal ones and for the state
(\ref{psi_in_dk}) overlaps  $\langle n l |\psi_{\text{in}}\rangle$ have positive
and substantial values for $\sim \kappa$ transverse and $\sim \delta^{-1}$
longitudinal successive modes. Nevertheless, even in this case results for
diagonal and non-diagonal initial states are pretty close (for
other parameters ranges and initial states the discrepancy is even smaller).




Consider now the most general case when both $\hat{\rho}_{\text{in}}$ and the
observable $\hat{A}$ have non-diagonal elements in the IS basis.
In this case, the value
$A_{\text{in}}$ defined by Eq.\ (\ref{AInTherm}) is different from the actual initial
expectation value. Rather, it corresponds to  
the infinite-time average for evolution of IS.
In addition, the correction to Eq.\ (\ref{ArelaxF}) 
\begin{eqnarray}
 A_{\text{nd}A}=-4\mathrm{Re}\sum_{\vec{n}\ne \vec{n}'}
\frac{\langle\vec{n} |\hat{A}|\vec{n}' \rangle}{(E_{\vec{n}}-E_{\vec{n}'})^{2}}
\bigl[
\eta_{\vec{n}'\vec{n}\vec{n}\vec{n}}^{(2)}(E_{\vec{n}})
\langle\vec{n} |\hat{\rho}_{\text{in}}| \vec{n} \rangle
\nonumber
\\
-\eta_{\vec{n}\vec{n}'\vec{n}\vec{n}'}^{(2)}(E_{\vec{n}})
\langle\vec{n}' |\hat{\rho}_{\text{in}}| \vec{n} \rangle
-\eta_{\vec{n}'\vec{n}'\vec{n}\vec{n}}^{(2)}(E_{\vec{n}})
\langle\vec{n} |\hat{\rho}_{\text{in}}| \vec{n}' \rangle
\bigr]
\label{AndA}
\end{eqnarray}
can be derived in the same way as Eq.\ (\ref{Andrho}). This correction is small
if the operator $\hat{A}$ does not couple energy-neighboring states of the
integrable system (it is a characteristic property of observables which do not act 
on some degrees of freedom). 

\begin{figure}
\includegraphics*[width=3.4in]{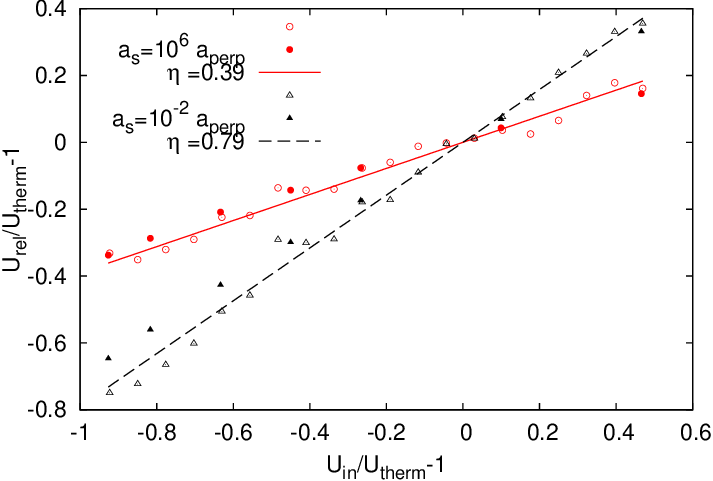}
\caption
{(Color online) The solid (red) and dashed (black) lines show predictions
(\protect\ref{ArelaxAv}) for expectation values of the transverse potential
energy $U$ for strong ($a_{s} = 10^{6} a_{\perp}$, $\bar{\eta}=0.39$) and weak
($a_{s} = 10^{-2} a_{\perp}$, $\bar{\eta}=0.79$) interactions, respectively, in
comparison to the infinite-time average (circles and triangles, respectively).
The filled and open symbols correspond to initial states
(\protect\ref{psi_in_dk}) and  (\protect\ref{psi_in_pn}), respectively. For all
the points the system energy is $E\approx 205 \hbar\omega_{\perp}$.
}
\label{Fig_U}
\end{figure}

As a concrete example of an observable we take the transverse potential energy
$\hat{U} = \mu\omega_{\perp}^2\rho^2/2$. This observable couples the states with
$l'=l$ and $n'=n,n\pm 1$.
Therefore $|E_{\vec{n}}-E_{\vec{n}'}|=2\hbar
\omega_{\perp }$ and the denominator in Eq.\ (\ref{AndA}) is large. 
Figure \ref{Fig_U} demonstrates good agreement between Eq.\ (\ref{ArelaxAv}) and
exact results. 
Both the initial state (\ref{psi_in_dk}) and the initial state 
 \begin{equation}
 \langle z,\rho |\psi_{\text{in}}\rangle 
\propto
\cos{\pi \zeta \over \delta}\theta \left(
 {\delta\over 2}-|\zeta |\right) \cos\left(2\pi l_{0}\zeta \right)
|n_{0}\rangle 
\label{psi_in_pn}
\end{equation} 
were used.
This state consists of two wavepackets moving in mutually opposite directions
along the waveguide axis (reminiscent of the experiment \cite{kinoshita2006}). 
Two completely different interaction strengths determine, with a good accuracy, 
the IPR values, which, for a given energy, are insensitive to the nature of the 
initial state . 




In conclusion, we demonstrate that an incompletely-chaotic non-integrable systems
with no selection rules relaxes to an equilibrium state which keeps a predictable 
memory of the initial state. 
We show that the
value after relaxation 
of a generic observable 
is given by a linear interpolation (\ref{ArelaxAv}) between the
thermal expectation value and the prediction 
of the ``diagonal ensemble'' for the underlying
integrable system (\ref{AInTherm}). 
The variable of this interpolation---the IPR (\ref{IPR})---is universal:
it is the same for all system observables 
and insensitive to the shape of the initial state. 
Microcanonical prediction further averaged over
the initial state should be used as the thermal expectation value
(\ref{AInTherm}).
This prediction is in a good agreement with the exact results on the relaxation in
a system of two zero-range-interacting atoms in a circular, transversely  harmonic
waveguide.



%
%

This work was supported by a grant from the Office of Naval Research ({\it
N00014-09-1-0502}) and a NSF grant ({\it PHY 1019197}).
%
%
%

\end{document}